\documentclass[aps,prd,multicol,reprint,showpacs,superscriptaddress,nofootinbib,amsfonts,amssymb,amsmath]{revtex4-1}  % for double-spaced preprint
\usepackage[english]{babel}
\usepackage[utf8]{inputenc}
\usepackage{graphicx}  % needed for figures
\usepackage{dcolumn}   % needed for some tables
\usepackage{bm}        % for math
\usepackage{amssymb}   % for math
\usepackage{setspace}
\usepackage[hidelinks]{hyperref}
\usepackage{verbatim}
\usepackage{indentfirst}
\usepackage{textcomp}
\usepackage{braket}
\usepackage{amsmath}
\usepackage{enumitem}
\usepackage{blindtext}
\usepackage{xcolor}
\usepackage{amsfonts}
\begin{document}
\author{Giovanni Montani}
\affiliation{Department of Physics, University of Rome “La Sapienza”, P.le Aldo Moro 5, Roma, 00185, Italy}
\affiliation{ENEA C. R. Frascati, Via E. Fermi 45, Frascati, Roma 00044, Italy}
%opening
\title{A Scenario for a Singularity-free Generic Cosmological Solution
}
\author{Roberta Chiovoloni}
\affiliation{Department of Physics, University of Swansea, Sketty, Swansea SA2 8PQ, UK}

\begin{abstract}
We develop a scenario for the emergence of a non-singular generic cosmological solution based on the a WKB characterization of one of the two anisotropy degrees of freedom.
We investigate the dynamics of the so-called inhomogeneous mixmaster in the ``corner'' configuration and inferring that one of the two anisotropic variables becomes small enough to explore the uncertainty principle.

Then, we apply a standard WKB approximation to the dynamics of the Universe which has macroscopic volume, one macroscopic anisotropy and one microscopic quantum degree of freedom. 

Our study demonstrates the possibility that the Universe acquires a non-singular classical behavior, retaining the quantum degree of freedom as a small oscillating ripple on a stationary Universe.
The role of the so-called ``fragmentation process'' is also taken into
account in outlining the generality of
such a behavior in independent local space regions.
\end{abstract}

\maketitle

\section{Introduction}

One of the most important contribution of the Landau School to theoretical
cosmology consisted of the dynamical characterization of the generic
cosmological solution in the vicinity of the primordial singularity
\cite{Belinsky:1970ew,Belinsky:1982pk,Montani:2011zz}. 
These studies, together with the general
theorem derived by Hawking and Penrose
\cite{Hawking:1969sw}, allowed to understand that the presence of a singular point in the past of our actual Universe should be regarded
as a general feature of the Einsteinian cosmology, not induced by the high
symmetry of the isotropic Universe geometry.

In 1963, Khalatanikov and Lifshitz published a paper \cite{Lifshitz:1963ps} which,
apart from the Lifshitz investigations on the isotropic Universe stability, contains a relevant analysis about relativistic cosmology in  a general framework.

In particular, %there
they derived the so-called ``generalized Kasner solution", i.e. the inhomogeneous extension of the Kasner solution,
describing the dynamics of the Bianchi I model \cite{Kasner:1921zz, LL, Montani:2011zz}.
They concluded that the asymptotic behavior of a generic inhomogeneous Universe toward the singularity is Kasner-like. However this conclusion was only partially correct. 
In fact, if on one hand the
generalized Kasner solution could possess the right number of four physically independent space functions, required for dealing in vacuum  in the general case, on the other hand, in order to survive up to the initial singularity, an inhomogeneous Kasner regime needs the imposition of an additional restriction, therefore loosing its general character.

The subsequent studies in \cite{Belinsky:1970ew} (for a detailed review see
\cite{{Montani:2007vu, Montani:2011zz}}), about the Bianchi VIII and IX models, clarified
how the asymptotic regime to the singularity requires an infinite sequence
of Kasner regimes (called Kasner epochs), which parameters are related by
a map having stochastic properties.
This picture was translated into an Hamiltonian formulation by Misner in
\cite{Misner:1969hg}. 
For a detailed discussion of the link existing between the Belinskii-Khalatanikov-Lifshitz (BKL) map between two Kasner
epochs and the Hamiltonian formulation in the so-called Misner-Chitr\'e -like variables, see \cite{Kirillov:1996rd}.
Misner called this Hamiltonian formulation of the original oscillatory regime, presented in \cite{Belinsky:1970ew}, the ``Mixmaster Universe'' (for a covariant
characterization of the Mixmaster chaos see \cite{Imponente:2001fy}).

This idea of an infinite sequence of Kasner regimes toward the cosmological
singularity was then implemented to the asymptotic dynamics of a generic
inhomogeneous model in \cite{Belinsky:1982pk}, see also \cite{Montani:2011zz} for a detailed re-analysis of this scenario.
This work completed the investigation in \cite{Lifshitz:1963ps}, by precising the
original statement: the generic inhomogeneous cosmological solution
approaches the cosmological singularity via an infinite sequence of Kasner regimes related, point by point in space, via a stochastic map.

However, in \cite{Belinsky:1982pk}, the inhomogeneous dynamics was described
assuming the existence of a single relevant spatial scale of inhomogeneity and the standard time evolution, associated to the
oscillatory regime, was recovered on a smaller spatial scale, roughly
identified with the average horizon size. 
However in \cite{KK1986} and \cite{Montani:1999aa} it was shown that the coupling between the space and time dependence of the metric tensor implies that smaller and smaller inhomogeneous scales are generated approaching the singularity, see \cite{Belinskii92} for a discussion of the impact that such a phenomenon can have on the primordial Universe turbulences, see also \cite{Barrow:2020rsp}. 

In \cite{Kirillov:1993aa} it was
demonstrated (see \cite{Montani:2011zz} for a simplified discussion) that the spatial gradients growth can not destroy the standard oscillatory regime because they grow slowly (in a logarithmic way) with respect to the terms which induce the instability of a Kasner regime and the transition to a new one.

However, more recent studies, see \cite{spike1, spike2}, demonstrated, mainly on a
numerical ground, the emergence of real spikes in the spatial gradients,
which put doubts on the nature of the generic inhomogeneous Mixmaster, as
the inhomogeneous extension of the oscillatory regime is commonly dubbed.

On how to reconcile the generic Mixmaster Universe with the highly
symmetric isotropic model, at least on a local spatial regime, see
\cite{Kirillov:2002kc} where the role of an inflationary regime is modeled via the effect of a massless scalar field plus a cosmological constant.

All these studies seem however to claim that the cosmological singularity is clearly present in the generic inhomogeneous solution, as described in the Einsteinian picture. 
Canonical quantum gravity in the metric approach seems unable to significantly change this situation, see the original work of Misner \cite{Misner:1969ae} or the more
recent analysis in \cite{Benini:2006xu}. 
The situation is different if the canonical quantization scheme is reformulated in Loop Quantum cosmology, see for instance the discussion in \cite{Ashtekar:2011ck}. 
A singularity-free generic cosmological
solution has been constructed in \cite{Antonini:2018gdd}, where the semi-classical Polymer dynamics (to be thought as the quasi-classical behavior of Loop Quantum Cosmology) is considered for the evolution toward the singularity. \\
For other approaches in extended theories of gravity, able to induce a bounce cosmology, see \cite{Bombacigno:2018tyw}, \cite{Parthasarathy_2020} and \cite{article}. 

\vspace{5pt} 
Here, the possibility for a singularity-free inhomogeneous Mixmaster is
based on a different scenario, in which the behavior of the Universe
during the so-called ``long era'' is examined. \footnote{The ``long era'' is intended as the Kasner era that the Universe undergoes when it is in the corner configuration.}
We investigate the possibility that, when such a configuration is addressed (according to the analysis in \cite{Lifshitz1970AsymptoticAO}), one of the two anisotropic degrees
of freedom is small enough to approach a quantum behavior since it can
explore the uncertainty principle in its own phase space. Then, we apply
the WKB scenario proposed in \cite{Vilenkin:1988yd} and we demonstrate that the
resulting Universe is a classical non-singular one, plus a small oscillating quantum anisotropy.

Using the language of the standard Hamiltonian formulation we outline
how, when the Universe performs a long era in the corner of the potential
term, a separation takes place between classical macroscopic components of
the inhomogeneous Mixmaster and a small quantum subset, made up of one of
the two anisotropic degrees of freedom.

%The possibility of such a scenario is suggested, on one hand, by the necessity, demonstrated in \cite{Lifshitz1970AsymptoticAO}, that at least one time the Universe must approaches the deep corner configuration in a given space point and, on the other
%one, on the so-called "fragmentation process'' discussed in \cite{Montani:1999aa}, allowing the repetition in space of such a situation.

The assessment of this scenario will relay on two main well-known results: 
\begin{enumerate}[label=(\roman*)]
\item \label{1} In \cite{Lifshitz1970AsymptoticAO}, it has been shown that, studying the statistical
properties of the BKL map, there always exists a significant probability
that the parameter $u$, characterizing a Kasner regime, acquires values large enough for the system dynamics to be deeply trapped in a corner configuration (where one of the two anisotropy degrees of freedom is very
small). 
\item \label{2} The existence of the fragmentation process discussed in
\cite{Montani:1999aa}. 
According to this, rational values taken by the function $u(x^i)$ across space can not be excluded from the evolution of
the BKL map, therefore even few steps of the BKL map ensure the existence of large values of $u$ in the neighborhood of certain space surface. 
\end{enumerate}

We observe how, while the result \ref{1} has a statistical character, being associated to the asymptotic iteration of the BKL map, the
result in \ref{2} can also be guaranteed by a finite deterministic implementation
of the BKL map across the space.

Our analysis is developed toward the singularity, but we can consider a
time reversed picture which is able to connect the standard inhomogeneous
Mixmaster to a primordial non-singular generic solution as soon as the
small quantum anisotropy degree of freedom is able to become a classical
variable, i.e. as soon as the Universe escapes the corner.

\section{Inhomogeneous Mixmaster}

In the ADM formalism, the line element of a generic inhomogeneous cosmological model, described by Misner variables $\alpha$, $\beta _+$ and $\beta_-$, reads as: 

\begin{equation}
	ds^2 = N^2 dt^2 - h_{ij}\left(dx^i + N^idt\right)\left( dx^j + N^jdt\right)
\, , 
	\label{ro1}
\end{equation}

with
\begin{align}
h_{ij} &= e^{2\alpha}\left(e^{2\beta}\right)_{ab}\ell^a_i\l^b_j \nonumber \\	
\beta &= diag \{\beta _+ + \sqrt{3}\beta_- , \beta _+ - \sqrt{3}\beta_- , -2\beta_+\}
\label{ro2}
\end{align}

Here, $N$ denotes the lapse function 
and $N^i$ the shift vector 
(these, together with the Misner variables, are space-time functions), while 
the vectors $\vec{l}^a$ ($a=1,2,3$) 
are linearly independent and they have generic space-dependent components. 
It has been shown in \cite{Belinsky:1982pk} 
that the time dependence of the vectors 
$\vec{l}^a$ is dynamically of higher order and it is associated with their 
rotation in space.
In the following we assume $8\pi G =1$.

The action associated to this generic 
model takes the following Hamiltonian representation \cite{Kirillov:1993aa, Montani:2011zz}:

\begin{equation}
	S_G = \int dtd^3x\left\{ 
	p_{\alpha}\partial _t\alpha + p_+\partial_t\beta _+ + p_-\partial_t\beta_- - N\mathcal{H} - N^i\mathcal{H}_i\right\},
	\label{ro3}
\end{equation}

$p_{\alpha}$, $p_+$ and $p_-$ 
are the conjugate momenta to $\alpha$, 
$\beta_+$ and $\beta _-$, respectively.
The super-Hamiltonian $\mathcal{H}$ admits 
the simplified expression:

\begin{equation}
	\mathcal{H} = \frac{1}{12}e^{-3\alpha} 
	\left\{ -p_{\alpha}^2 + p_+^2 + p_-^2 + e^{4\alpha}V_G(\beta _+,\beta_-)\right\}
	\, , 
	\label{ro4}
\end{equation}

where the potential term $V_G$ is obtained neglecting all the spatial gradients of the Misner variables in the spatial curvature.
On a classical level, this approximation is justified \emph{a posteriori} by demonstrating that such gradients increase slowly towards the singularity with respect to the time derivatives of the configurational variables \cite{Kirillov:1993aa}.
This scenario leads to the so-called inhomogeneous Mixmaster model, i.e. within each smooth spatial scale (roughly the horizon scale), the dynamics is isomorphic to that one of 
a homogeneous Mixmaster \cite{Misner:1969ae, Belinsky:1970ew}. 
However in \cite{KK1986, Kirillov:1993aa, Montani:1999aa}, it has been shown that, in the inhomogeneous Mixmaster, the chaotic time evolution couples to the spatial dependence and increasingly small scales are generated for the space variation of the Misner variables, but without destroying the dynamical scheme 
of infinite sequence of Kasner regimes. 

The classical dynamics of a generic cosmological models is described by the Hamilton equations associated to the Misner variables and by the constraints 
obtained variating the action $S_G$ respect to $N$ and $N^i$, namely:

\begin{equation}
\mathcal{H} = \mathcal{H}_i = 0 
\, .
	\label{ro4.1}
\end{equation}

In the inhomogeneous Mixmaster approximation, the super-momentum constraint 
reduces to the following dominant contribution:

\begin{equation}
	p_{\alpha}\partial _i\alpha + 
	p_+\partial _i\beta_+ + p_-\partial _i\beta_- = 0
	\, . 
	\label{ro5}
\end{equation}

This constraint is consistent with the 
scalar nature of the Misner variables under reparametrization of the spatial coordinates, which acts on the vectors 
$\vec{l}^a$ only. 

We conclude this dynamical picture
assigning the explicit expression for the potential term $V_G$, namely:

\begin{align}
&V_G = \frac{1}{4}\Big(C_1^2e^{4\beta_++4\sqrt{3}\beta_-} + C_2^2e^{4\beta_+-4\sqrt{3}\beta_-} + C_3^2e^{-8\beta_+}\Big) \nonumber \\
&- \frac{1}{2}\Big(C_1C_2e^{4\beta_+} - C_1C_3e^{-2\beta_++2\sqrt{3}\beta_-} - C_2C_3e^{-2\beta_+-2\sqrt{3}\beta_-}\Big)
\label{ro6}
\end{align}

Above, the generic functions $C_a(x^i)$
defining the inhomogeneity character of the cosmological model, can be expressed via the vectors $\vec{l}^a$ as 
$C_a \equiv \vec{l}^a\cdot rot \vec{l}^a$ (expressions to be intended in Euclidean sense with respect to the coordinates $x^i$ and the vector components $l^a_i$).

\begin{figure}[h!]
\includegraphics[scale=0.35]{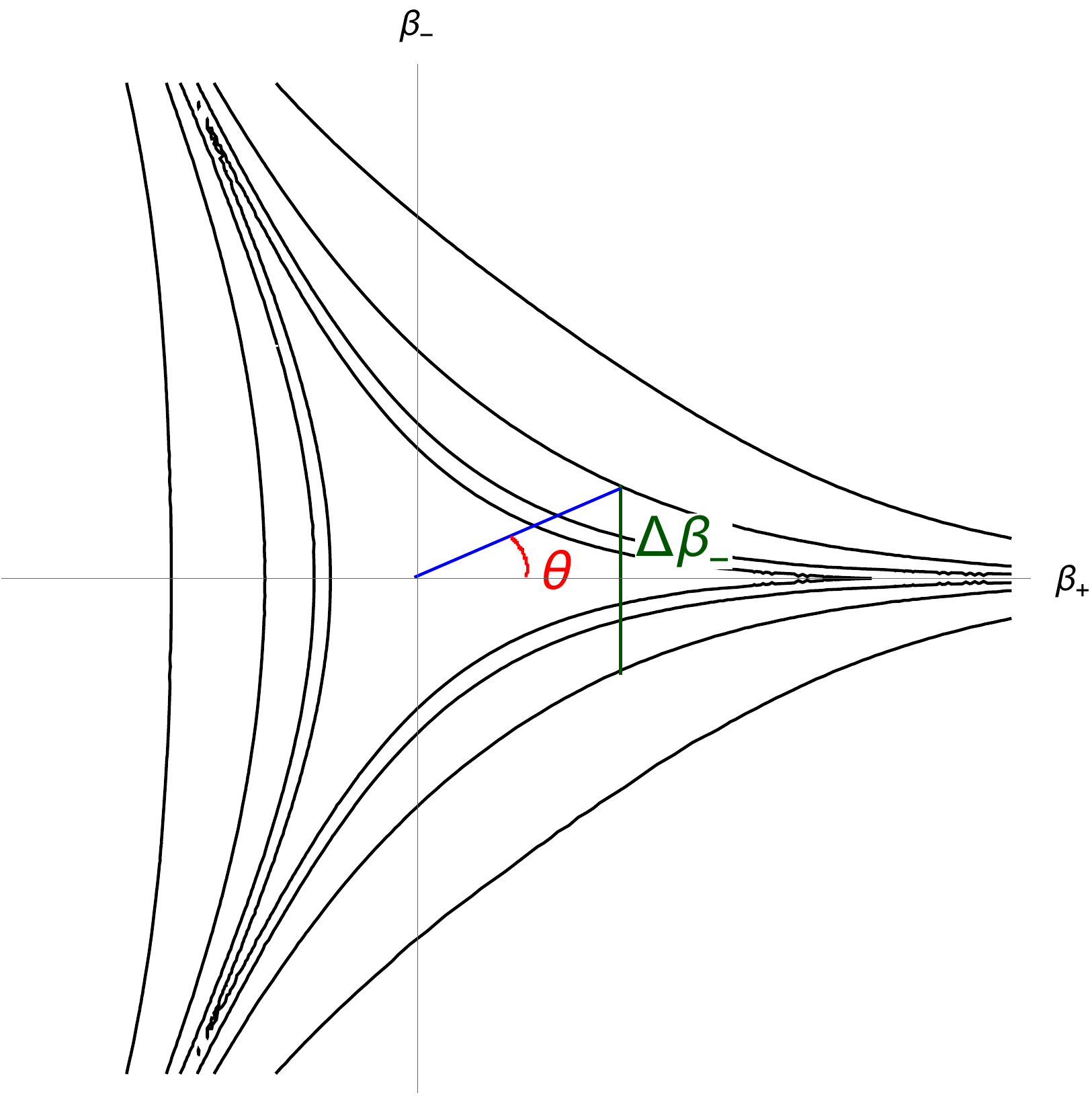}
\caption{Equilateral triangle formed by the equipotential lines in each space point. 
The segment $\Delta \beta_-$ is in green, while the angle $\theta$ is represented in red.}
\label{Bianchi_IX}
\end{figure}

The equipotential lines associated to this potential form, in each space point a curvilinear equilateral triangle (Fig.\ref{Bianchi_IX}), having three open corners reaching infinity. Here, we will focus our analysis on the system dynamics when the interior of one of these corners is considered.

 %and since, they are equivalent to each% other (it is sufficient to rotate by $\pi/3$ the  
%coordinate plane $\{\beta_+,\beta_-\}$ to map one into another), we consider the corner along the axis 
%$\beta_-= 0$. 
The three corners are equivalent, this can be shown simply rotating the coordinate plane $\{\beta_+,\beta_-\}$ by $\pi/3$ to map one corner into another. Therefore without loss of generality we consider the corner along the axis 
$\beta_-= 0$. 

From a geometrical point of view, the corner configuration corresponds to deal with 
two space directions scaling essentially with the same oscillating time law, while the remaining one decays monotonically  toward the singularity.
It is worth noticing that, on a classical level, an inhomogeoneous Mixmaster scheme is well-established \cite{Kirillov:1993aa,Montani:1999aa,Belinskii92}, apart from the emergence of spikes in the spatial gradients  \cite{spike1, spike2}, while, on a quantum level, it stands as an ansatz to be validated \emph{a posteriori} and it is commonly referred 
as the BKL Conjecture.

\subsection{Generalized Kasner solution}

If the initial singularity is identified with the instant of time when the spatial volume of the Universe (i.e. the three-metric determinant) vanishes, then we can fix that singularity with the limiting value $\alpha \rightarrow -\infty$. 

In such a limit the potential $V_G$ tends to become an infinite well, in which center $\beta_+\sim \beta_-\sim 0$ (actually for an increasing region as the singularity is approached) the potential term can be neglected and the generic inhomogeneous  Universe is described by the so-called \emph{generalized Kasner solution} 
\cite{Lifshitz:1963ps}. 

It is immediate to recognize that, when $V_G$ is negligible, the momenta $p_{\alpha}$ and the two $p_{\pm}$ are all constant in time and the following relations, obtained making use of the first Hamilton equations, hold

\begin{equation}
	\frac{d\beta_{\pm}}{d\alpha}=\frac{p_{\pm}}{p_{\alpha}}\equiv \pi_{\pm}(x^i)\, ,\, \Rightarrow \beta_{\pm}=\pi_{\pm}(x^i)\alpha + \bar{\beta}_{\pm}(x^i) 	\, ,
	\label{ro12}
\end{equation}

where $\bar{\beta}_{\pm}$ denote generic space functions.

Since the functions $\pi_{\pm}$ 
must satisfy, by definition, the relation $\pi_+^2 + \pi_-^2 = 1$, we can set 
$\pi_+ = \cos \theta$ and $\pi _-=\sin \theta$. The function $\theta (x^i)$ 
changes at each bounce against the potential walls and it acquires a random behavior. 
Therefore the system can reach a configuration deeply in the corner $\beta_-=0$, if $\sin \theta \simeq \theta \simeq \epsilon \ll 1$.

\section{Quantum small oscillations}

Let us now investigate more in detail the structure of the generic inhomogeneous model Hamiltonian in the corner configuration. 

If we choose a space coordinate system $\bar{x}^i$, such that 
$C_1(\bar{x}^i) = C_2(\bar{x}^i) \equiv C(\bar{x}^i)$,
then, the super-Hamiltonian constraint reads, inside the corner 
($\{ \beta_+\gg 1\, ,\, \beta_-\ll 1\}$), as
\begin{equation}
	- p_{\alpha}^2 + p_+^2 
	+ \mathcal{H}^- = 0 
	\, , 
	\label{rox20}
\end{equation}

where $\mathcal{H}^-$ is a small contribution and it is defined as:

\begin{equation}
	\mathcal{H}^- \equiv p_-^2 + 
	6 C^2e^{4(\alpha + \beta_+)}\beta_-^2
	\, ,
	\label{rox21}
\end{equation}

At a fixed $\alpha$ value, the coordinate interval for the variable 
$\beta_-$ in the corner is of the order $\Delta \beta_- \sim  2\beta_+ \theta = \beta_+\epsilon$. as shown in Fig.\ref{Bianchi_IX}. 
Furthermore, according to the generalized Kasner solution (i.e. comparing the kinetic and the potential term in $\mathcal{H}^-$), we get 

\begin{equation}
	\frac{\mathcal{H}^-}{p_{\alpha}^2}\sim 
\frac{\mathcal{H}^-}{p_+^2} \sim \pi_-^2 \sim \theta ^2 = \epsilon^2
	\, .
	\label{rox22}
\end{equation}

where, in the first part we used that, from the super-Hamiltonian constraint applied to the semi-classical Misner variables, $p^2_{\alpha} = p^2_+$. 

If the BKL map generates a small value of $\epsilon$ of order $\sqrt{\hbar}$ (here we disregard the physical dimensions of $\hbar$ to avoid the use of two small parameters, one physical and one dimesionless), then the Hamiltonian constraint (\ref{rox20}) can be decoupled, according to the analysis in \cite{Vilenkin:1988yd}, into a classical part, 
associated to the variables $\alpha$ and $\beta_+$ plus a quantum small subsystem, constituted by the anisotropy degree of freedom $\beta_-$, which lives on the space-time defined by the classical components. 
By other words, we are inferring that the variable $\beta_-$ is enough small to explore the uncertainty principle with $\Delta \beta_-\le 2\sqrt{\hbar}\beta_+$ and $\Delta p_-\ge 2\sqrt{\hbar}/\beta_+$. The quantum subsystem 
shows to possess the ``smallness'' requirement postulated in \cite{Vilenkin:1988yd} and precised in \cite{Agostini:2017oql}.
Under the hypotheses above, the Universe state functional can be written as follows

\begin{equation}
	\Psi = \exp \{ i\Sigma (\alpha ,\beta_+)/\hbar\}\Phi (\alpha , \beta_+, \beta_-)
	\, , 	\label{rox23}
\end{equation}

where $\Sigma$ is associated to 
the classical system, while $\Phi$ describes the quantum subcomponent.
According to the scheme developed by 
\cite{Vilenkin:1988yd}, the functional 
derivative of $\Phi$ with respect to 
the space field $\beta_-(x^i)$ are of order of $1/\sqrt{\hbar}$, therefore $\mathcal{H}^-\Phi \propto \mathcal{O}(\hbar)$. 

To obtain the dynamical implications of 
the state function (\ref{rox23}), we need to apply the canonical operator version of the constraint (\ref{rox20}) and of the super-mometum constraint \eqref{ro5} i.e.:

\begin{equation}
	\left[ \hbar^2\frac{\delta^2}{\delta \alpha^2} - \hbar^2\frac{\delta^2}{\delta \beta_+^2} + \hat{\mathcal{H}}^-\right] \Psi = 0
	\, .
	\label{rox24}
\end{equation}
\begin{equation}
i\hbar \left( \partial_i\alpha \frac{\delta \,}{\delta\alpha} +
\partial_i\beta_+\frac{\delta\,}{\delta \beta_+} + \partial_i\beta_-
\frac{\delta\,}{\delta \beta_-}\right)\Psi = 0
\, .
\label{super-mom_constraint}
\end{equation}
where the symbol $\delta$ denotes functional derivatives.

%Also the operator version of the %supermomentum constraint must be %implemented, namely $%\hat{\mathcal{H}}_i\Psi = 0$.

At the zero approximation order in $\hbar$ we get  the classical Hamilton-Jacobi super-Hamiltonian and super-momentum equations for the variables $\alpha$ and $\beta_+$, i.e. the following system of functional differential equations

\begin{eqnarray}
-\left( \frac{\delta \Sigma}{\delta \alpha}\right)^2 + \left(\frac{\delta\Sigma}{\delta \beta_+}\right)^2 = 0
	\label{rox25a}\\
	\frac{\delta \Sigma}{\delta\alpha}\partial _i\alpha + \frac{\delta \Sigma}{\delta \beta_+}\partial _i\beta_+ = 0 \, .
	\label{rox25b}
\end{eqnarray}

In other words, the classical component is associated to the reduced action 

\begin{align}
S_{Class} = \int dtd^3x \{ p_{\alpha}\partial_t\alpha + &p_+\partial_t\beta_+ - \frac{N}{12}e^{-3\alpha}\left( -p_{\alpha}^2 + p_+^2\right)\nonumber \\
& - N^i\left( p_{\alpha}\partial_i\alpha + p_+\partial_i\beta_+\right)\} 
\label{rox26}
\end{align}

By a simple algebra, it is possible to show that the quantum functional $\Phi$ obeys the equation

\begin{equation}
	i\hbar \partial_t \Phi = \left\{\int d^3x \frac{N}{12}e^{-3\alpha} \left( \hat{\mathcal{H}}^-\Phi + i\hbar\left(\frac{\delta^2 \Sigma}{\delta \alpha^2} - \frac{\delta^2 \Sigma}{\delta\beta_+^2}\right) \Phi\right)\right\}
\, , 
	\label{rox27}
\end{equation}

where 

\begin{equation}
	\partial_t\Phi \equiv \int d^3x \left\{\left( \partial _t\alpha \frac{\delta \,}{\delta \alpha} + \partial_t\beta_+ 
	\frac{\delta \,}{\delta \beta_+}\right)\Phi\right\}
	\, , 
	\label{rox28}
\end{equation}

$\partial_t\alpha$ and $\partial_t\beta_+$ being calculated from the action 
(\ref{rox26}) and via the identification of the momenta with the corresponding functional derivatives of $\Sigma$. 
\\
To derive \eqref{rox27}, we made also use of the semi-classical part of
the order $\hbar$ of the super-momentum constraint \eqref{super-mom_constraint}, i.e.

\begin{equation}
\partial_i\alpha \frac{\delta \Phi}{\delta \alpha} + \partial_i\beta_+
\frac{\delta \Phi}{\delta \beta_+} = 0
\, ,
\label{nuova19}
\end{equation}

which states the invariance of the wave functional $\Phi$ with respect to
the space coordinates in the classical
line element. 
\\

The present analysis differs from the 
approach presented in \cite{Vilenkin:1988yd} (see also \cite{Battisti:2009qd, DeAngelis:2020wjp,Chiovoloni:2020bmh}) because we are dealing with a functional formalism, due to the inhomogeneity of the considered model, and we are taking the variables $\alpha$ and $\beta_+$ as strictly classica. This last difference results in 
the last term in parentheses of Eq. 
(\ref{rox27}).

It is immediate to check that 
Eq. (\ref{rox25a}) admits the following solution: 

\begin{equation}
	\Sigma = \int d^3x K(x^i)\left( \alpha + \beta_+\right) 
	\, , 
	\label{rox29}
\end{equation}

which, according to the Hamilton-Jacobi method, yields the classical relation 

\begin{equation}
	\alpha + \beta_+ = \beta_0(x^i)
	\, . 
	\label{rox30}
\end{equation}
which, substituted in \eqref{rox21}, leads to
\begin{equation}
\label{H_beta0}
\hat{\mathcal{H}}^- = p^2_- + 6 C^2 e^{4\beta_0}\beta_-^2. 
\end{equation}
In order for the solution (\ref{rox29}) to satisfy the super-momentum equation (\ref{rox25b}), it is enough to require $\beta_0=const.$ 

From the classical action 
(\ref{rox26}), we recognize the following relation between 
the variable $\alpha$ and the synchronous time $T$:

\begin{equation}
	\alpha = \frac{1}{3}\ln \frac{T}{T_0}
	\, , 
	\label{rox31}
\end{equation}

where $T_0$ is a generic instant.

Choosing, without loss of generality, the vector $\vec{l}^3$ along the coordinate $x^3$, the classical solution above makes the 
line element (\ref{ro1}) of the form

\begin{equation}
	ds^2 = dT^2 -\left( 
	\frac{T}{T_0}\right)^2e^{-4\beta_0} 
	\left(dx^3\right)^2 - \left(dl_2\right)^2  
	\label{rox32}
\end{equation}

where $(dl_2)^2$ is a static two-dimensional line element on the plane $\{x^1,x^2\}$.
As well-known \cite{Belinsky:1970ew,Belinsky:1982pk}, 
the line element above is 
associated to a non-singular cosmological model and it becomes 
static as soon as we pass to 
new coordinates $T^{\prime}=(T/T_0) cosh x^3$ and $x^{3^{\prime}}=(T/T_0)sinh x^3$. 

Using the expression (\ref{rox29}) of $\Sigma$ and introducing the time variable $\tau$ defined via the lapse function 
$N=12e^{3\alpha}$, the quantum functional equation (\ref{rox27}) 
reduces to the form

\begin{equation}
	i\hbar\partial_{\tau}\Phi = \int d^3x\hat{\mathcal{H}}^-\Phi
	\, , 
	\label{rox33}
\end{equation}

with $\hat{\mathcal{H}}^-$ defined in $\eqref{H_beta0}$. 

The dynamical decoupling of the space points, i.e. of each space region sufficiently smooth 
(so that spatial gradients are negligible), allows to reduce the Superspace to the collection of local Minisuperspace, each for each point $x^i$. Thus, we can write:

\begin{equation}
	\Phi = \Pi_{x^i}\phi _{xi}(\tau , \beta_-)
	\, ,
	\label{rox34}
\end{equation}

where the local wave functions 
$\phi_{x^i}$ satisfy the equations

\begin{equation}
	i\hbar\partial_{\tau}\phi_{x^i} = \left\{ -\hbar^2\partial_{\beta_-}^2 + 6C^2(x^i)e^{4\beta_0} \beta^2_-\right\} \phi_{x^i}
	\, .
	\label{rox35}
\end{equation}

The functional $\Phi$ must also 
satisfy the quantum component of the supermomentum constraint, 
i.e.:

\begin{equation}
	-i\hbar \frac{\delta \Phi}{\delta \beta_-}\partial_i\beta_- = 0 
	\, . 
	\label{rox36}
\end{equation}

However, when we take the functional $\Phi$ in the factorized form (\ref{rox34}), we are inferring that it is naturally satisfying Eq. (\ref{rox36}), simply because that approximation corresponds to deal locally with the 
condition $\partial_i\beta_-\simeq 0$.
Here, we are implementing the BKL conjecture, based on the idea that the scale of spatial gradients is larger than the quantum correlation length. In this sense, we are re-introducing the concept of ``quantum causality'': space regions that evolve independently are not in causal contact. 

Eqs (\ref{rox34}) have the morphology of quantum harmonic oscillators each in each space point and it is well-known that localized non-spreading states can be always constructed. We expect that the variable $\beta_-$ can be represented by localized state because when it enters the corner is a classical degree of freedom and its available domain remains of order $\hbar\beta_+$ in that configuration.

Thus, we can conclude that, if our scheme is reliably applicable to the Universe dynamics deeply entering the corner, the cosmological singularity is removed 
because we get a classical non-singular space-time on which very small quantum fluctuations of the variable $\beta_-$ live.
Such an intriguing picture is well-established when it is referred to a given spatial point (causal region), but to understand how it works in the Universe as a whole, we need to develop some 
considerations on the BKL map \cite{Belinsky:1970ew,Belinsky:1982pk} and on the so-called ``fragmentation process'' 
\cite{Montani:1999aa} (see also \cite{KK1986}). 

We conclude this section by emphasizing that the picture proposed above can be reversed in time and we could start with a non-singular classical Universe with a small quantum anisotropy and, 
as the space volume increases (i.e. $\alpha$ increases), this degree of freedom becomes classical, so that the dynamics comes out of the corner configuration and
the full configurational domain is restored. By other words, in this scenario, the generic inhomogeneous cosmological solution can emerge from a non-singular initial configuration and then evolves toward the standard oscillatory regime discussed in \cite{Belinsky:1970ew,Belinsky:1982pk}.

\section{Inhomogeneous BKL map}

If we introduce the parameter $u(x^i)$ \cite{Belinsky:1982pk} the quantities $\pi_+(x^i)$ and $\pi_-(x^i)$ take the explicit form:

\begin{equation}
	\pi_+ = \frac{ u^2 + u - 1/2}{u^2 + u + 1}\, ,\, 
	\pi_- = \frac{\sqrt{3}}{2}\frac{2u + 1}{1 + u + u^2}
	\, .
	\label{rox37}
\end{equation}

In the present context, we can restrict these expressions to $ u \gg 1$ since the considered corner corresponds to very large values of $u$ (note that $\theta\simeq\sqrt{3}/u$). 
To understand when such large values appear, we have to consider the BKL map \cite{Belinsky:1970ew} which provides the value $u^{\prime}$ generated from the initial value $u$ via the effect of the potential wall in the standard oscillatory regime, i.e.: 

\begin{equation}
	for \, u > 1 \, u^{\prime}=u-1 \, ,\, for \, u\le 1 \, u^{\prime}=\frac{1}{u-1}
	\, . 
	\label{rox38}
\end{equation}

All the initial rational values of the parameter $u=u^0$ are evolved for a finite number of the BKL map steps, after which the value $u\rightarrow \infty$ (i.e. $\theta=0$) is recovered. 
Instead, the irrational values of the initial parameter $u^0$ evolve indefinitely and the BKL map outlines a strong (exponential) instability with respect to the initial condition: 
if we modify the value $u^0$ by a very small amount, the sequence of values generated by the map iteration becomes uncorrelated with respect to the sequence 
associated to $u^0$, after few steps. 
We stress that the rational values of $u$ were excluded in the original analysis in \cite{Belinsky:1970ew,Belinsky:1982pk}, because, being of zero measure on the real axis, they turn to be a non-general 
initial condition.
However, if we assign, over the inhomogeneous space, the initial condition $u=u^0(x^i)$, the rational values can not clearly be excluded simply for continuity reasons. Thus, each spatial region containing surfaces on which $u$ is rational, enters deeply the corner, after a certain number of iterations of the map and our scenario can be implemented 
close enough to one of such regions. 

Actually, when the parameter $u$ is thought as a physical parameter, we have to assign its values with a given uncertainty, 
even because the Kasner solution to which it is associated is an approximate regime obtained by neglecting the potential walls. 
This consideration, together with the instability of the BKL map, leads to think of $u$ as a statistically distributed variable and it can be shown that it admits the following steady probability density \cite{Montani:1999aa} 

\begin{equation}
	w(u) = \frac{1}{\ln 2}
	\frac{1}{u\left(u + 1\right)} 
	\, . 
	\label{rox39}
\end{equation}
In \cite{Lifshitz1970AsymptoticAO}, it has been shown that, starting from a generic initial value $u^0$, the situation of a very large $u$ is always reached, at least one time, as the BKL map is iterated for sufficiently long time. Actually, the BKL map has, especially when expressed in terms of the
fractional part of the parameter $u$
see \cite{Barrow:1981sx}, "strong mixing'' properties and
therefore, starting from a generic
irrational value of $U$ all the other irrational ones (including very
large values) are, soon or later, generated. 
This result ensures that, also from a statistical point of view, in each point of the space 
(enough smooth space region), 
the conditions for the system to deeply entering the corner are reached. 

However, in \cite{Montani:1999aa} it has been argued how the iteration of the BKL map in two close space points gives uncorrelated values of the parameter $u$ after some steps and thus is at the ground of the progressive increasing of the spatial gradients towards the initial singularity. 
As a consequence of this result, the proposed scenario takes place in different instant of time in dynamical independent space regions. Nevertheless, once the system enters the corner, the BKL map is no longer applicable, because two potential walls are simultaneously relevant. 
Furthermore, once our paradigm is implemented, the increasing behavior of the spatial gradients is naturally stopped. 
Each smooth space region is 
characterized by a non-singular static space-time and the statistical properties of the BKL map are reflected only on the specific initial condition at which the corner dynamics is implemented.

\section{Conclusions}

We investigated the possibility to obtain a non-singular generic cosmological solution as result of a quantum behavior of the small anisotropy $\beta_-$ within a deep corner configuration.
In other words, we separated the Universe dynamics into a classical non-singular one, plus a quantum effect which manifests in a simple small oscillation of $\beta_-$ according to a time-independent frequency.

In order to establish this configuration, we inferred that, for a sufficiently large value of the parameter $u$,
the variable $\beta_-$ is extremely small well-inside the corner of the
potential, so that it explores the uncertainty principle.

To characterize the generality of the proposed scheme, we made use of two
complementary effects: 
\begin{enumerate}[label=(\roman*)]
\item In each assigned space point, the iteration of the BKL map is associated to a significant probability for very long era, i.e. a trapping of the system dynamics deeply in the corner, see
\cite{Lifshitz1970AsymptoticAO}. 
\item The existence of the so-called fragmentation process,
i.e. the impossibility to exclude rational values of $u$ in a continuous
function representation $u(x^i)$, which generates on all the corresponding
space surfaces exactly the limit $\beta_-\equiv 0$, with associated
neighborhoods where a long era must take place \cite{Montani:1999aa}.
\end{enumerate}

This analysis completes and generalizes the consideration made in
\cite{Chiovoloni:2020bmh} about the
WKB approach to the homogeneous case, see also \cite{Battisti:2009qd,
DeAngelis:2020wjp} for related topics. 
The basic motivation for such a generalization consists in the natural
character that the corner configuration acquires in the inhomogeneous
picture, as effect of the fragmentation process. This means that few iterations of
the  map can be enough to generate very high values of $u$ in correspondence of all the rational values of the initially assigned
function $u^0(x^i)$.

In \cite{Barrow:2020dap} it has been argued the possibility for a
synchronization of the dynamics of different spatial regions of the
inhomogeneous mixmaster. Without entering in the discussion of such a
proposal and its validity, we observe that such a synchronization
would reduce the relevance of the spatial gradients, in favor of an
homogeneous-like picture. The proposed feature would likely reduce the
impact of the fragmentation process, but would not prevent the realization of the present scenario, according to the point i) above.

The transition of the
inhomogeneous mixmaster to a new regime of gravitational turbulence
could instead be of different impact on the present scenario, as
inferred in \cite{Belinskii92}, see
also \cite{Barrow:2020rsp}. In this case it would be clear the applicability of the potential
representation in a fully turbulent Universe.
Finally, about the possible implications of the rotation of the vectors $\vec{l}^a$ in the presence of a matter source, like a perfect fluid
(a question not yet fully explored in the inhomogeneous sector), see
\cite{Belinski:2014kba}.

\bibliography{Mont2}
\bibliographystyle{ieeetr}

\end{document}